\documentclass[11pt]{article}
\usepackage{moriond,epsfig}
\usepackage{amsmath,amssymb,hyperref,url,xspace}

\def\be{\begin{equation}}
\def\ee{\end{equation}}
\def\bea{\begin{eqnarray}}
\def\eea{\end{eqnarray}}

\newcommand {\dphi}     {\ensuremath{\Delta\phi}}
\newcommand {\pt}       {\ensuremath{p_T}}
\newcommand {\AJ}       {\ensuremath{A_J}}
\newcommand {\npart}    {\ensuremath{N_{\rm part}}}

\newcommand {\pp}    {\mbox{pp}}

\newcommand {\PbPb}  {\mbox{PbPb}}

\newcommand{\GeVc}{\ensuremath{\text{GeV/}c}\xspace}

\begin{document}

%
% Cover
%

\begin{figure}[htbp!]
\begin{center}
\includegraphics[width=0.99\textwidth]{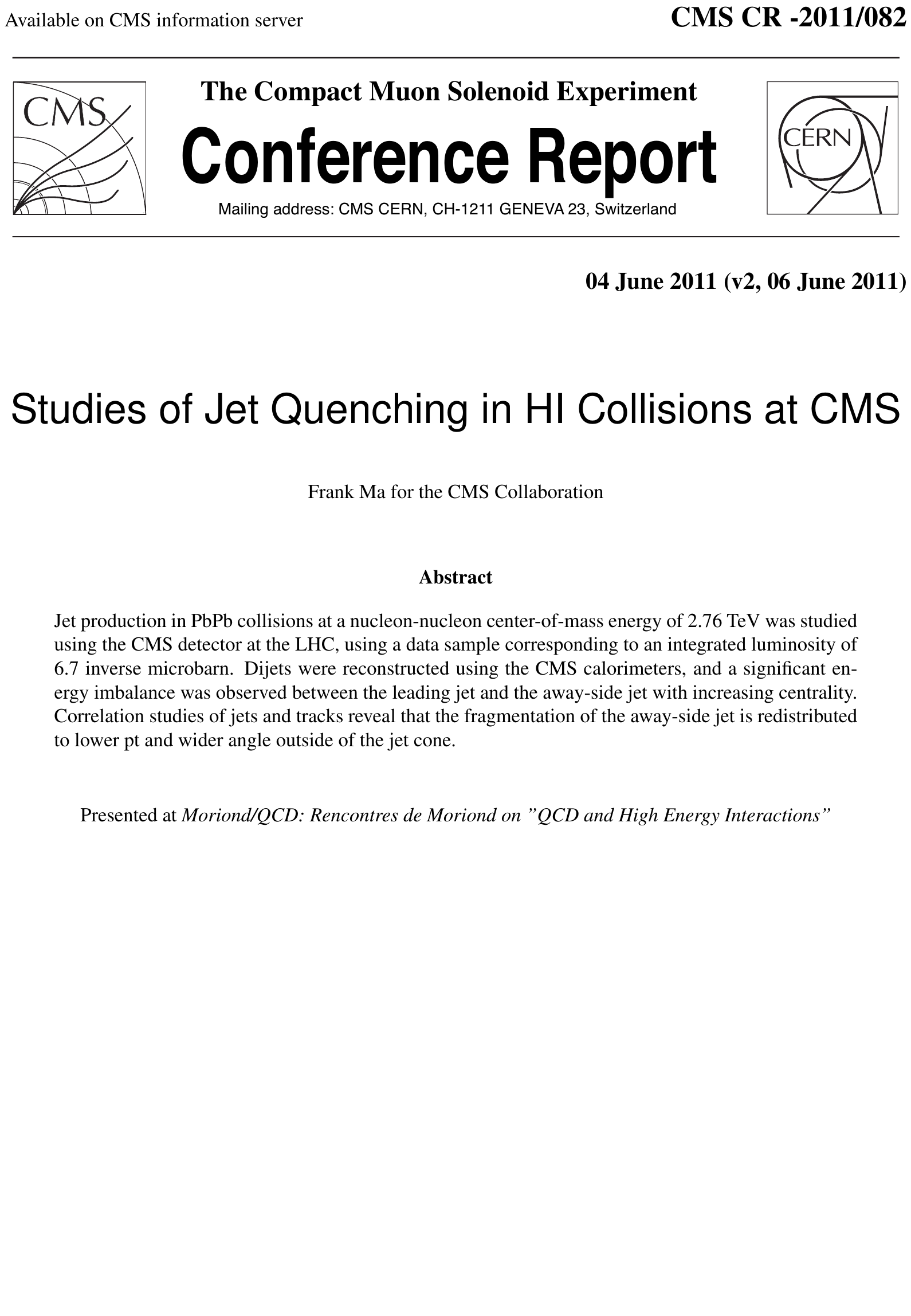}
\label{fig:Cover}                  
\end{center}
\end{figure}

\vspace*{4cm}
\title{Studies of Jet Quenching in HI Collisions at CMS}

\author{ Frank Ma for the CMS Collaboration }

\address{Massachusetts Institute of Technology\\
Cambridge, Massachusetts 20139, USA}

\maketitle\abstracts{
Jet production in PbPb collisions at a nucleon-nucleon center-of-mass energy of 2.76 TeV was studied using the CMS detector at the LHC, using a data sample corresponding to an integrated luminosity of 6.7 inverse microbarn. Dijets were reconstructed using the CMS calorimeters, and a significant energy imbalance was observed between the leading jet and the away-side jet with increasing centrality. Correlation studies of jets and tracks reveal that the energy of the away-side jet is redistributed to lower pt and wider angle outside of the jet cone.}

\section{Introduction}
% Big Picture
Heavy ion collisions at the Large Hadron Collider (LHC) allow one to study the thermodynamic properties of the fundamental theory of the strong interaction --- Quantum Chromodynamics (QCD).
% Phenomenology
Studying the modification of jets that are created from within the medium has long been proposed as a particularly useful tool for probing the QCD medium properties~\cite{Appel:1985dq,Blaizot:1986ma}.
In the presence of a QCD medium, the partons may lose energy to the medium via elastic processes (collisional parton energy loss) or inelastic processes (radiative parton energy loss).
The study of medium-induced modifications of dijet properties can
therefore shed light on the transport properties of collective QCD matter created by heavy ion collisions.

\section{Experimental Methods}
% Dataset
This analysis was performed using the data collected in 2010 from  PbPb collisions at a nucleon-nucleon center-of-mass energy of $\sqrt{s_{_{NN}}}=2.76$~TeV at the Compact Muon Solenoid (CMS) detector~\cite{bib_CMS}.
Jets were reconstructed with background subtraction based on their energy deposits in the CMS calorimeters~\cite{Kodolova:2007hd}, and the events were selected from a jet-triggered dataset~\cite{CmsJetQuenching:2011}.

% Centrality
Because heavy ions are extended objects, the impact parameter is an important characterization of the events.
The amount of overlap between the two colliding nuclei is what we mean by ``centrality'' of the collision.
In this analysis, centrality was determined from minimum events based on the total energy from both  forward hadronic calorimeters~\cite{CmsJetQuenching:2011}.
Simulations can be used to correlate centrality, as quantified using the fraction of the total interaction cross section, with physically meaningful quantities such as the total number of nucleons in the two
lead ($^{208}$Pb) nuclei which experienced at least one inelastic collision (\npart).

\section{Results}

\subsection{Dijet Properties in pp and PbPb data}
\label{sec:dijet_results}
To obtain a clean dijet selection, we select events with a leading 
jet having corrected $p_{\mathrm{T},1} > 120$~\GeVc, a subleading jet with $p_{\mathrm{T},2} > 50$~\GeVc,  and a minimum azimuthal angle between them ($\Delta \phi_{12} > 2\pi/3$). Only jets within $|\eta| < 2$ were considered.
Given this selection, we observe a sharp $\Delta \phi_{12}$ correlation between leading and subleading jets~\cite{CmsJetQuenching:2011} , indicating true dijet pairs.

% Introduce Aj
In-medium induced parton energy loss can significantly alter the detector level jet energy (and hence dijet energy balance) by either transporting energy outside of the jet cone or shifting the energy towards low
momentum particles that will not be detected in the calorimeter.
To characterize the dijet momentum balance quantitatively, we use the asymmetry ratio,
\begin{equation}
\label{eq:aj} 
A_J = \frac{p_{\mathrm{T},1}-p_{\mathrm{T},2}}{p_{\mathrm{T},1}+p_{\mathrm{T},2}}~,
\end{equation} 
where $\pt$ is the corrected $\pt$ of the reconstructed calorimeter jet. The subscript $1$ always refers to the leading jet, so that $A_J$ is positive by construction.
%The use of an energy ratio in $A_J$ allows overall uncertainties in the jet energy scale to cancel.

\begin{figure}[ht!]
\begin{center}
\includegraphics[width=0.60\textwidth]{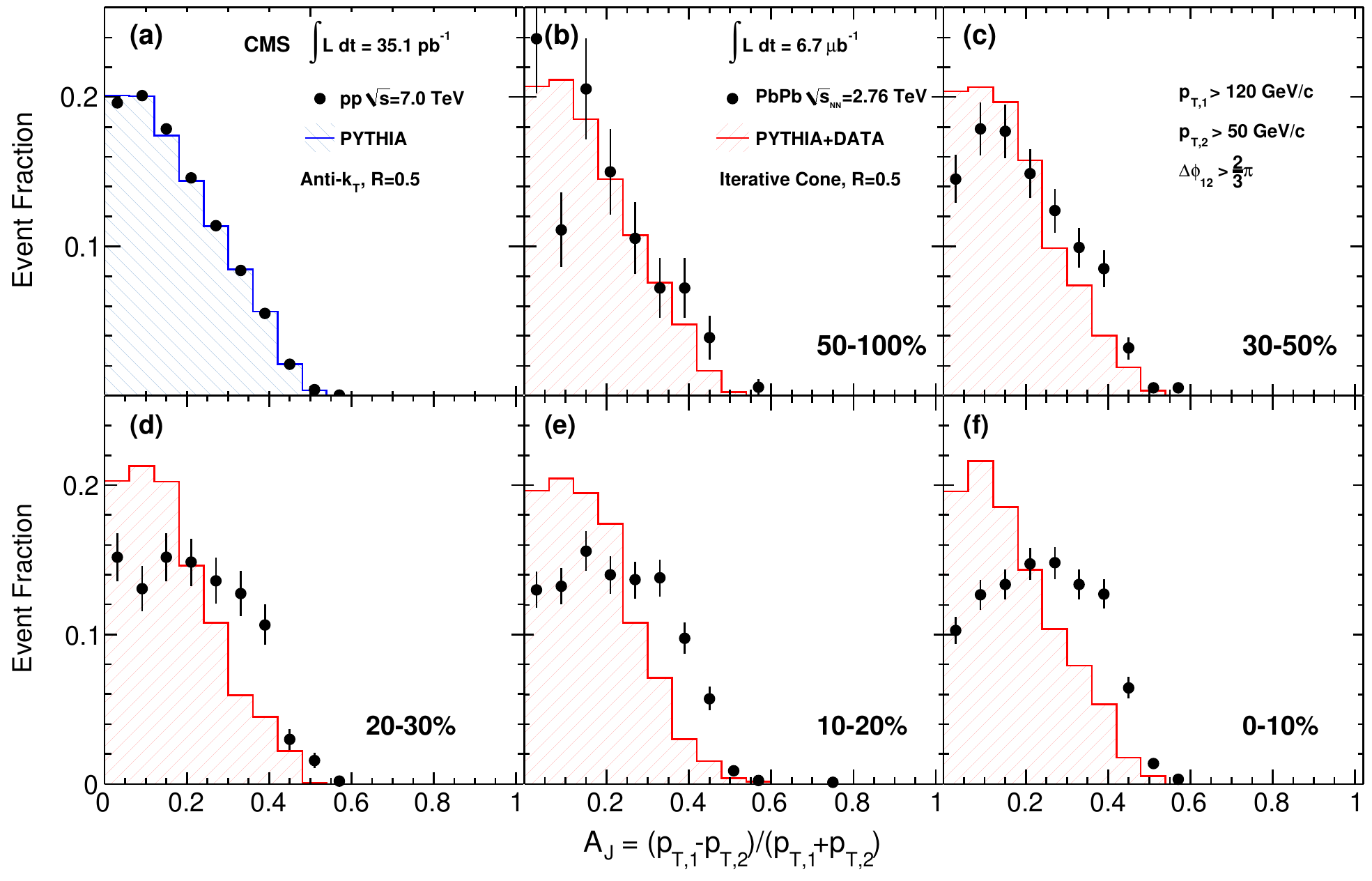}
\includegraphics[width=0.39\textwidth]{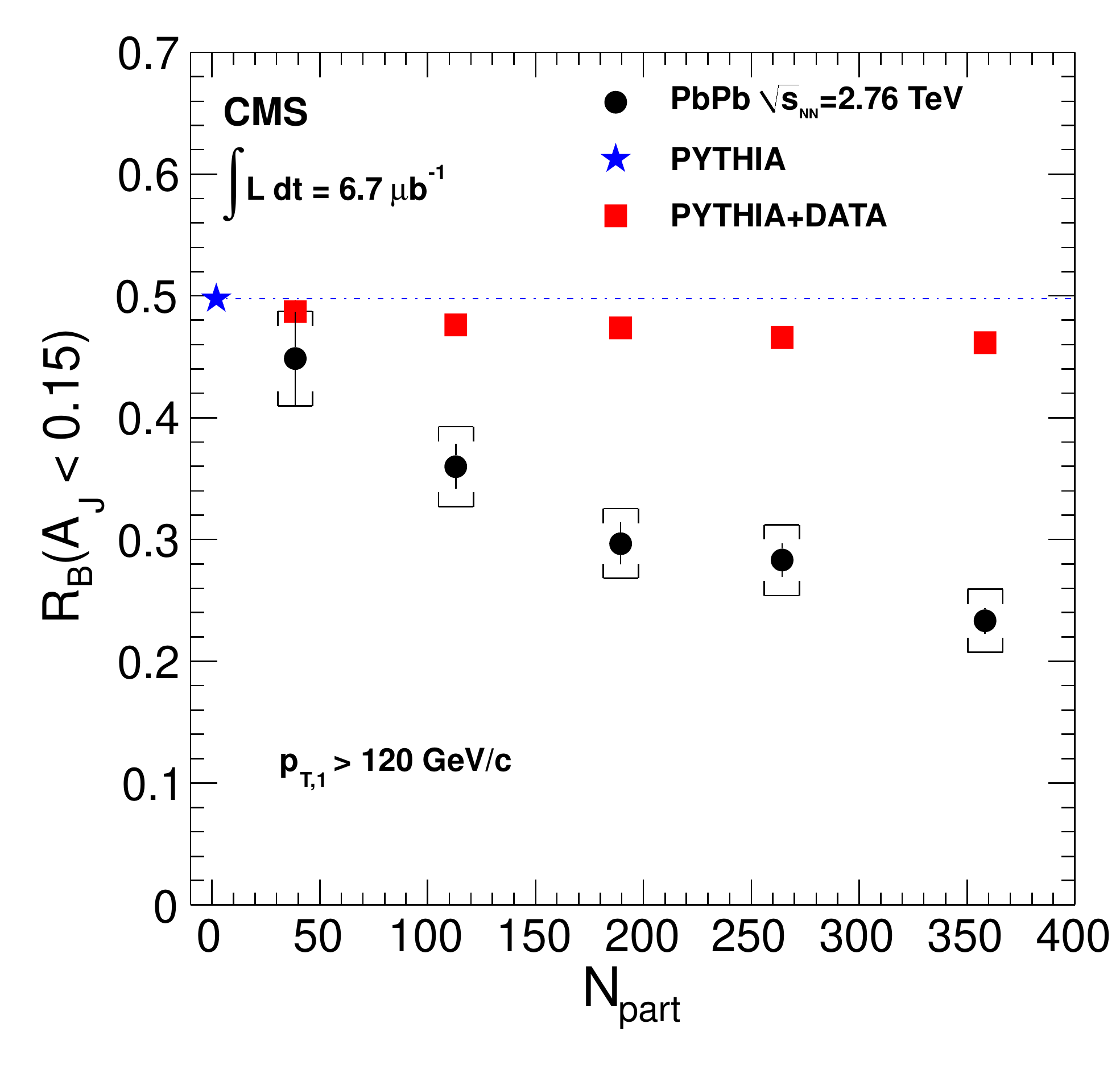}
\caption{Left 6 panels show dijet asymmetry distribution, $A_{J}$, of selected dijets  for 7 TeV pp collisions (a) and 2.76 TeV \PbPb\ collisions in several centrality bins:  (b) 50--100\%, (c) 30--50\%, (d) 20--30\%, (e) 10--20\% and (f) 0--10\%.
Data are shown as black points, while the histograms show  (a) {\sc{pythia}} events and  (b)-(f) {\sc{pythia}} events embedded into \PbPb\ data.  
Right panel shows fraction of selected dijets with $A_J<0.15$ out of all events with a
leading jet with $p_{\mathrm{T},1} > 120$~\GeVc\, as a function of \npart.
The result for reconstructed {\sc{pythia}} dijet events (blue filled star)
is plotted at \npart\ = 2.  The other points (from left to right) correspond to centrality bins shown in (b)-(f) in the left 6 panels. 
The red squares are for reconstruction of {\sc{pythia+data}} events and the filled
circles are for the \PbPb\ data.
For the data points, vertical bars and brackets represent the statistical and systematic uncertainties, respectively.}
\label{fig:JetAsymm}                  
\end{center}
\end{figure}

In Fig.~\ref{fig:JetAsymm} (a),  the $A_J$ dijet asymmetry observable calculated 
by {\sc{pythia}} is compared to \pp\  data at $\sqrt{s}$ = 7~TeV.
%The {\sc{pythia}} dijet events were processed with the full detector simulation and analysis chain.
We see that data and event generator are found to be in excellent agreement, demonstrating that {\sc{pythia}}
(at $\sqrt{s}$ = 2.76~TeV) can serve as a good reference for the dijet imbalance analysis
in \PbPb\ collisions.
Figs.~\ref{fig:JetAsymm}~(b)-(f) show the centrality dependence of $A_J$ for \PbPb\ collisions.
To separate effects due to the medium itself from effects simply due to reconstructing jets
in the complicated environment of the underlying PbPb event, the reference {\sc{pythia}} dijet events were embedded into a minimum bias selection of PbPb events at the raw data level~\cite{CmsJetQuenching:2011}.
In contrast to {\sc{pythia+data}}, we see that data shows a dramatic decrease of balanced dijets with increasing centrality.

The centrality evolution of the dijet momentum balance can be explored 
more quantitatively by studying the fraction of balanced jets in the \PbPb\ events.
The balanced fraction, $R_B(\AJ < 0.15)$, is plotted as a function of collision centrality (in terms of \npart) in 
the right panel of Fig.~\ref{fig:JetAsymm}. It is defined as the fraction of all events with a leading jet having $p_{\mathrm{T},1} > 120$~\GeVc\ for which a subleading partner with $\AJ < 0.15$ and $\dphi_{12} > 2\pi/3$
is found.
The \AJ\ threshold of 0.15 was chosen because it is the median of the \AJ\ distribution
for selected dijets in pure {\sc{pythia}} events.
In contrast to {\sc{pythia+data}}  dijets, the \PbPb\ data show a rapid 
decrease in the fraction of balanced jets with collision centrality.
The effect is much larger than the combined systematic uncertainties.
These results imply a degradation of the parton energy, or jet quenching, in the medium produced in central \PbPb\ collisions.
The final systematic uncertainties, stemming mainly from uncertainties in the jet energy scale, are described in ~\cite{CmsJetQuenching:2011}.

\subsection{Overall Momentum balance of Dijet Events}
\label{sec:missingpt}
% Introduce Met
We next turn to the question of where and how the medium energy loss occurs by exploiting additional information from the entire CMS tracker.
We measure overall transverse momentum balance in the dijet events using the projection of missing $\pt$ of reconstructed charged tracks onto the leading jet axis, defined as,
%\begin{eqnarray*}
\begin{equation}
\displaystyle{\not} p_{\mathrm{T}}^{\parallel} = 
-\sum_{\rm i}{p_{\mathrm{T}}^{\rm i}\cos{(\phi_{\rm i}-\phi_{\rm Leading\ Jet})}},
%\end{eqnarray*}
\end{equation}
where the sum is over all tracks with $\pt > 0.5$~$\GeVc$ and $|\eta| < 2.4$.
%The results were then averaged over events to obtain $\langle \displaystyle{\not} p_{\mathrm{T}}^{\parallel} \rangle$.
For this study, the leading and subleading jets are required to have a slightly smaller $\eta$ range ($|\eta| < 1.6$)
to allow the jets to remain fully inside the CMS tracker acceptance.
No background subtraction in the track distribution is needed since the underlying \PbPb\ tracks cancel in the $\displaystyle{\not} p_{\mathrm{T}}^{\parallel}$ sum.

\begin{figure}[ht!]
  \begin{center}
    \includegraphics[width=0.7\textwidth]{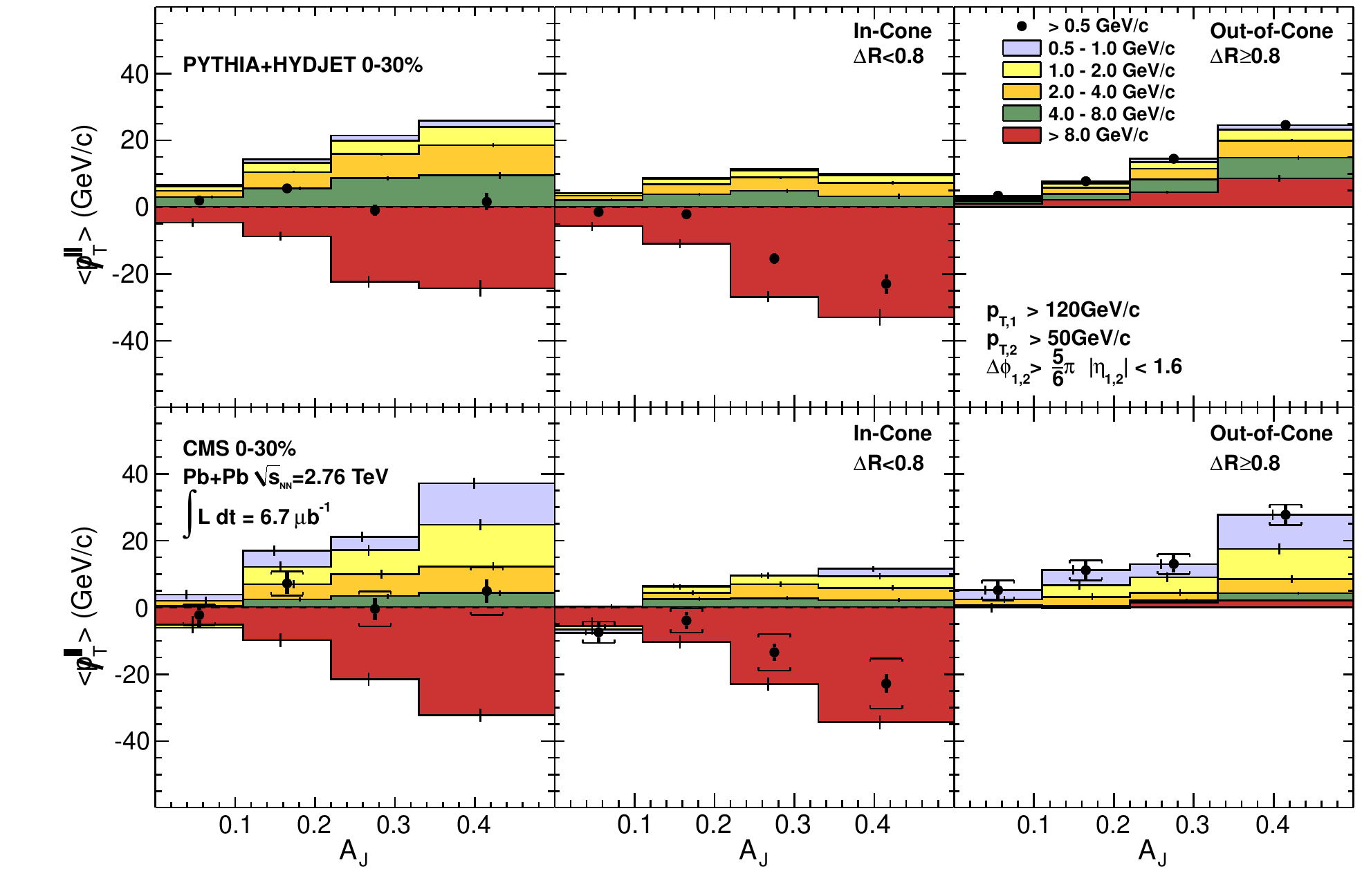}
    \caption{Average missing transverse momentum, 
$\langle \displaystyle{\not} p_{\mathrm{T}}^{\parallel} \rangle$, 
for tracks with $\pt > 0.5$~\GeVc, projected onto the leading jet axis (solid circles).
The $\langle \displaystyle{\not} p_{\mathrm{T}}^{\parallel} \rangle$ values are shown as a function of dijet asymmetry $A_J$ in 0--30\% central events, for the full event (left), inside ($\Delta R < 0.8$) one of the leading or subleading jet cones (middle) and outside ($\Delta R > 0.8$) the leading and subleading jet cones (right).
For the solid circles, vertical bars and brackets represent 
the statistical and systematic uncertainties, respectively.
Colored bands show the contribution to $\langle \displaystyle{\not} p_{\mathrm{T}}^{\parallel} \rangle$ for five
ranges of track \pt. The top and bottom rows show results for {\sc{pythia+hydjet}} and \PbPb\ data, respectively. 
For the individual $\pt$ ranges, the statistical uncertainties are shown as vertical bars. Note that as the underlying \PbPb\ event in both data and MC is not $\phi$-symmetric on an event-by-event basis, the back-to-back requirement was tightened to $\dphi_{12} > 5 \pi/6$.
}
    \label{fig:MissingpT}
  \end{center}
\end{figure}

In Fig.~\ref{fig:MissingpT}, $\langle \displaystyle{\not} p_{\mathrm{T}}^{\parallel} \rangle$
is shown as a function of \AJ\  in the 0--30\% centrality bin, where we expect the medium effects to be the strongest. Here \AJ\ is the same calorimeter jet \AJ\ as described in Sec.~\ref{sec:dijet_results}.
The left column shows $\langle \displaystyle{\not} p_{\mathrm{T}}^{\parallel} \rangle$ using all selected tracks. One sees that in both data and simulation, the overall momentum balance of the events (shown as solid circles) is recovered within uncertainties even for dijets with large energy asymmetry. 
This cross-checks the soundness of the detector, since regardless of medium effects, net transverse momentum is conserved. The figure also shows the contributions to 
$\langle \displaystyle{\not} p_{\mathrm{T}}^{\parallel} \rangle$ for five transverse momentum ranges from 0.5--1~\GeVc\ to $\pt > 8$~\GeVc, shown as stacked histograms.

Important insights into the dijet asymmetry emerge when we look at the $\langle \displaystyle{\not} p_{\mathrm{T}}^{\parallel} \rangle$ differential in radial distance from the jets.
The middle and right columns of Fig.~\ref{fig:MissingpT} show $\langle \displaystyle{\not} p_{\mathrm{T}}^{\parallel} \rangle$ separately for tracks inside cones of size $\Delta R = 0.8$ around the leading and subleading jet axes, and for tracks outside of these cones.
We see that for both data and MC an in-cone imbalance of $\langle \displaystyle{\not} p_{\mathrm{T}}^{\parallel} \rangle \approx
-20$~\GeVc\ is found for the $\AJ > 0.33$ selection.
This shows that track momentum sums within the leading and subleading jet cones confirm the calorimeter dijet asymmetry results showed earlier in Sec.~\ref{sec:dijet_results}.
In addition, both data and simulation show similar large negative contribution to $\langle \displaystyle{\not} p_{\mathrm{T}}^{\parallel} \rangle$ (i.e., in the direction of the leading jet) in the $\pt > 8$~\GeVc\ range.
This cross-checks that the dijet energy asymmetry in data is not caused by fake jets from background fluctuation, because only genuine high $\pt$ jets give rise to high $\pt$ tracks.
Looking now at the right column, we see that in both data and MC the in-cone energy difference is balanced by a corresponding out-of-cone 
imbalance of  $\langle \displaystyle{\not} p_{\mathrm{T}}^{\parallel} \rangle \approx 20$~\GeVc. However, in the \PbPb\ data 
the out-of-cone contribution is carried almost entirely  by tracks with $0.5 < \pt < 4$~\GeVc\, whereas in MC more than 50\% of the balance 
is carried by tracks with $\pt > 4$~\GeVc, with a negligible contribution from $\pt < 1$~\GeVc. 
The {\sc{pythia+hydjet}} results are indicative of semi-hard initial or final-state radiation as the underlying cause for large \AJ\ events in the MC study.
This is in contrast to the results for large-\AJ\ \PbPb\ data, which show that a
large part of the momentum balance is carried by soft particles ($\pt < 2$~\GeVc) and radiated
at large angles to the jet axes ($\Delta R > 0.8$).

\section{Summary and Conclusion}
A strong increase in the fraction of highly unbalanced jets has been
seen in central PbPb collisions compared with peripheral collisions and
model calculations, consistent with a high degree of parton energy loss in the produced QCD medium.  A large fraction of the momentum balance of these unbalanced jets is 
carried by low-\pt particles at large radial distance, in contrast
to {\sc{pythia}} simulations embedded into heavy ion events.
The results provide qualitative constraints on the nature of the jet modification 
in \PbPb\ collisions and quantitative input to models of the transport properties of the medium
created in these collisions.

%%%%%%%%%%%%%%%%%%%%%%%%%%%%%%%%%%%%%%%%%%
\section*{References}
\providecommand{\href}[2]{#2}

\end{document}